\documentclass[aps,prd,showpacs,twocolumn,superscriptaddress,preprintnumbers,10pt,floatfix]{revtex4}

\usepackage{amssymb}  
\usepackage{graphicx}
\usepackage{aas_macros}
\usepackage[dvips]{color}

\newcommand{\gsim}{\, \raisebox{-0.8ex}{$\stackrel{\textstyle >}{\sim}$ }}

\newcommand{\beq}{\begin{equation}}
\newcommand{\eeq}{\end{equation}}
\newcommand{\ba}{\begin{array}}
\newcommand{\ea}{\end{array}}
\newcommand{\bea}{\begin{eqnarray}}
\newcommand{\eea}{\end{eqnarray}}
\newcommand{\bi}{\begin{itemize}}  
\newcommand{\ei}{\end{itemize}}
\newcommand{\ben}{\begin{enumerate}} 
\newcommand{\een}{\end{enumerate}}
\newcommand{\bc}{\begin{center}}
\newcommand{\ec}{\end{center}}

\usepackage[normalem]{ulem}  

\begin{document}
\title{Superfluid Heat Conduction and the Cooling of Magnetized Neutron Stars}
\author{Deborah N. Aguilera}
\affiliation{Tandar Laboratory, Comisi\'on Nacional de Energ\'ia At\'omica, Av. Gral. Paz 1499, 
1650 San Mart\'in, Buenos Aires, Argentina}
\author{Vincenzo Cirigliano}
\affiliation{Theoretical Division, Los Alamos National Laboratory,
 Los Alamos, New Mexico 87545, USA}  
\author{Jos\'e A. Pons}
\affiliation{Department of Applied Physics, University of Alicante, 
Apartado de Correos 99, E-03080 Alicante, Spain}
\author{ Sanjay Reddy}
\affiliation{Theoretical Division, Los Alamos National Laboratory,
 Los Alamos, New Mexico 87545, USA}  
\author{Rishi Sharma}
\affiliation{Theoretical Division, Los Alamos National Laboratory,
 Los Alamos, New Mexico 87545, USA}  

\begin{abstract}
We report on a new mechanism for heat conduction in the neutron star crust. We find that collective modes of superfluid neutron matter, called superfluid phonons (sPhs), can 
influence heat conduction in magnetized neutron stars. They can dominate the heat conduction transverse to magnetic field when the magnetic field $B \gsim 10^{13}$ G. At density $\rho \simeq 10^{12}-10^{14} $ g/cm$^3$ the conductivity due to sPhs is significantly
larger than that due to lattice phonons and is comparable to electron conductivity when temperature $\simeq 10^8$ K. This new mode of heat conduction can limit the surface anisotropy in highly magnetized neutron stars.  Cooling curves of magnetized neutron stars with and without superfluid heat conduction could show observationally discernible differences. 
\end{abstract}
\maketitle
Multiwavelength observations of thermal emission from the neutron star (NS) surface and explosive events such as superbursts in accreting NS and giant-flare in magnetars provide a real opportunity to probe the NS interior (see \cite{YakovlevPethick:2004} and \cite{Page:2006ud} for recent reviews). Theoretical models of these phenomena clearly underscore the importance of heat transport in the neutron star (NS) crust and have shown that it directly impacts observations. For example, nearby isolated compact X-ray sources that have also been detected in the optical band
have a significant optical excess relative to the extrapolated
X-ray blackbody emission (a factor 5 to 14) \cite{Haberl2007}. This optical excess can arise naturally if heat conduction in the NS crust is anisotropic due to the presence of a large magnetic field leading to an anisotropic surface temperature distribution \cite{Geppert2004,Pons:2005,Geppert2006}. In accreting NSs, the thermal conductivity directly affects the observed thermal relaxation time of the crust \cite{Shternin:2007}, and the superburst ignition depth and recurrence time-scale \cite{Brown:1998}.


Typically, heat conduction is thought to arise due to the flow of relativistic electrons. Electrons are numerous, degenerate, and provide an efficient mode to transport heat. Conduction due to lattice phonons has been considered, but shown to be unimportant when the crust temperature $T\ge10^6$ K  \cite{Pons:2005,Chugunov:2007} and in NSs with low magnetic fields. In this letter we demonstrate that a new mechanism for heat transport arising due to 
the superfluid nature of the inner crust is important at high temperature and/or large magnetic fields. Here the heat is carried by the collective excitations of the neutron superfluid called superfluid phonons (sPhs). 

This mode of conduction is especially relevant in NSs with moderate to high magnetic fields (B\gsim$10^{13}$ G).  In these cases, 
electron heat transport is very anisotropic because electrons can only move freely along 
magnetic field lines. Motion perpendicular to the field is restricted because the 
electron-cyclotron frequency is large compared to the inverse collision time. In contrast, sPhs, being electrically neutral excitations, are not directly affected by magnetic fields. We find that sPh conduction is typically much larger than the electron conduction transverse to the field and this limits the degree of  surface temperature anisotropy in magnetized NSs.  In NSs with low magnetic fields, sPh  conduction can still be relevant when the temperature is in the range $10^8-10^9 $ K.

Free neutrons in the inner crust are known to form Cooper pairs and become superfluid at 
temperatures below the critical temperature $T_c \approx 10^{10}$ K.  The superfluid ground state spontaneously breaks 
baryon-number symmetry and gives rise to a new massless Goldstone 
mode - the superfluid phonon (sPh). At long wavelengths, these phonons 
have a linear dispersion relation $\omega = v_s~ q $, 
where $v_s \simeq k_{\rm Fn}/ \sqrt{3} M$,  $M$ is the mass of the 
neutron and $k_{\rm Fn}$ is the neutron Fermi momentum. 
The thermal conductivity of a weakly interacting gas of sPhs can be 
computed from kinetic theory and is given by  
\begin{equation}  
\kappa_{\rm sPh}= \frac{1}{3}~C_V~v_s~\lambda_{\rm sPh}
\label{conduct}
\end{equation} 
where $C_V=  2\pi^2~T^3/(15~v_s^3)$  is the specific heat of the phonon gas and 
$\lambda_{\rm sPh}$ is the typical mean free path of a thermal sPh.  
We can rewrite Eq.~\ref{conduct} using fiducial values of the temperature and the 
phonon velocity as 
\begin{equation} 
\kappa_{\rm sPh}=1.5 \times 10^{22} ~\left( \frac{T}{10^8 \rm ~K}\right)^3
~\left(\frac{0.1}{v_s}\right)^2~\left(\frac{\lambda_{\rm sPh}}{\rm cm}\right)
~\frac{\rm erg}{\rm cm~s~K}\,. 
\label{conduct2}
\end{equation}
This is to be compared with the thermal conductivity of electrons in the inner crust 
which is approximately $10^{18}$ erg cm$^{-1}$ s$^{-1}$ K$^{-1}$ when T$=10^8$ K 
\cite{Shternin:2006}. In the presence of large magnetic fields electron conduction in 
the direction perpendicular to the field can be several orders of magnitude smaller. 
Although sPhs are less numerous than the electrons they have large mean free paths because they interact weakly at long-wavelengths. In what follows we will estimate $\lambda_{\rm sPh}$ and its temperature dependence.
%

The low energy degrees of freedom in the inner crust are lattice phonons (lPhs), electrons, and sPhs. At long-wavelengths, the excitations  of the ion-lattice with mean number density 
$n_I$, mass number $A$ and charge $Z$, are the longitudinal and transverse lPhs with velocities 
$c_s=\omega_P/q_{\rm TFe}$ and $c_t \simeq 0.7~ \omega_P/q_{\rm BZ}$, respectively \cite{FI76}.  
Here $\omega_P=\sqrt{4 \pi e^2 Z^2 n_I/(AM)}$  is the ion-plasma frequency, 
$q_{\rm TFe} = \sqrt{4 e^2/\pi} ~p_{\rm Fe}$ is the inverse Thomas-Fermi screening length of the 
electrons, $ p_{\rm Fe}$ is the electron Fermi momentum, $q_{\rm BZ}=(4.5 \pi)^{1/3}~a_i^{-1}$ 
and $a_i$ is the inter-ion distance. The electrons are characterized by their Fermi momentum $p_{\rm Fe} = (3 \pi^2 Z n_I)^{1/3} \gg m_e \gg T$ and form a highly degenerate, relativistic Fermi gas. As mentioned earlier, the superfluid phonons (sPhs) are collective excitations of the neutron superfluid and their interactions are weak at low temperature. Their scattering cross  section is parametrically suppressed by the factor  $ T^8/\mu_n^8$ where  $\mu_n= k_{\rm Fn}^2/2M$ is the neutron chemical potential  \cite{Son:2005rv}. We note that when $T\simeq T_c$,  sPhs are strongly damped as they can easily decay into neutron particle-hole excitations. However, since the crust cools to $T \ll T_c$ within several hours of the birth of the NS, this processes is exponentially suppressed by the factor $\exp{\left( -2 T_c/T\right)}$ and the mean free path of sPhs is limited only by their interaction with lPhs, electrons and impurities. 

\begin{figure}[h]
\begin{center}
\includegraphics[width=8.5cm, clip]{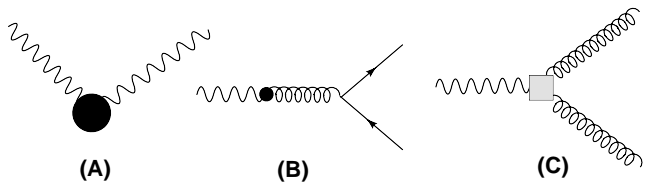}
\end{center}
\caption{Phonon scattering processes: (A) Rayleigh scattering; 
(B) phonon absorption by electrons; and (C) decay of sPhs into lattice phonons. 
The wavy-line is the sPh, the curly-line is the lPh, and the solid line are electrons, respectively.} 
\label{fig:scatter}
\end{figure}

The relevant processes are illustrated in Fig.~\ref{fig:scatter}:   (i) Rayleigh scattering of phonons due to interactions with (compositional) impurities 
in the solid lattice \cite{Baym:1967}; (ii) absorption of sPhs due to their mixing 
with the longitudinal lattice  phonons which are absorbed efficiently by electrons; 
and (iii) the decay of a sPh into two lattice phonons (lPhs). 

At long-wavelength, the cross section for Rayleigh scattering  of phonons by impurities is 
\begin{equation} 
\sigma_{\rm R} = \pi r_0^2~\frac{(q~r_0)^4}{1+(q~r_0)^4} \,,
\end{equation} 
where $q$ is the phonon momentum and $r_0$ is an intrinsic strong-interaction length-scale related to the scattering length of the neutron-impurity system. This depends on the impurity composition and is expected to be about 10 fm. At low temperature, when $q ~r_0 \ll 1$, the mean free path of sPhs with thermal energy $\omega_{\rm th} = 3 T$ is 
\begin{equation} 
\lambda_{\rm Ray}  = (n_{\rm Im} \sigma_{\rm R})^{-1} = \frac{ v_s^4}{81~n_{\rm Im}~ \pi~ r_0^6~ T^4} \,, 
\label{lambdaR1}
\end{equation} 
where $n_{\rm Im} = 3/(4\pi d^3)$ is the impurity number density and 
$d$ is the inter-impurity distance. Using fiducial values we can rewrite Eq.~\ref{lambdaR1} as  
\begin{equation} 
\lambda_{\rm Ray} = 450~\left(\frac{v_s}{0.1}\right)^4 \left(\frac{x}{10}\right)^3
\left(\frac{10 ~\rm fm}{r_0}\right)^3
~T_7^{-4}~{\rm cm} 
\label{lambdaR2}
\end{equation}
where $x=d/r_0$ is the diluteness parameter for the impurities and $T_7$ is the temperature
in $10^7$ K. 

We now show that inelastic processes such as sPh absorption by electrons and decay to lPhs  shown in Fig.~\ref{fig:scatter} (B) and (C) are more relevant.  
To study these processes we need a low-energy effective theory which couples 
the sPhs and lPhs.  On general grounds, this effective theory can only contain derivative 
terms and at low temperature it is sufficient to retain the leading terms. The leading 
order Lagrangian that describes these interactions is of the form  
\begin{eqnarray} 
{\cal L}_{\rm eff}&=& g_{\rm mix}~\partial_0 \phi ~\partial_i \xi^i +  \frac{g_{\rm mix}}{\Lambda^2} ~\partial_0 \phi ~\partial_i \xi_i  \partial_i \xi_i    \nonumber  \\
&+& \frac{g_{\rm mix}}{\Lambda_t^2}  ~~\partial_0 \phi ~\partial_i \xi_j \partial_i \xi_j + \cdots
\label{eqn:eftlag}
\end{eqnarray}
where $\phi$ and $\xi_{\rm i=1\dots 3}$ are the sPh and lPh (canonically normalized) fields, respectively.  
The first term describes the mixing between longitudinal lPhs and sPhs, and the second and third terms  
describe the process for the decay of a sPh into two longitudinal or transverse lPhs, respectively. 

The coefficients of this effective theory are determined by computing the coupling 
between neutrons and lattice phonons in a manner similar to the calculation of the electron-phonon 
coupling in condensed matter physics \cite{FetterWalecka}. We begin with the fundamental short-range interaction between neutrons and ions given by 
\begin{eqnarray}   
{\cal H}_{\rm nI}=\int d^3x ~ d^3x'~ \Psi_I^\dagger(x) \Psi_I(x)~V(x-x')~\psi_n^\dagger(x)\psi_n(x)\nonumber  
\end{eqnarray} 
where $V(x-x')= 2 \pi a_{\rm nI}~\delta^3(x-x')/M$, is the neutron-ion low-energy potential and $a_{\rm nI}$ is the neutron-ion scattering length. By expanding the ion and neutron density fluctuations in terms of their collective modes we obtain
\begin{equation}
g_{\rm mix} =2 ~a_{\rm nI}~\sqrt{\frac{n_{\rm I} ~k_{\rm F_n}}{A ~M^2}}\,, \quad
\Lambda^2 =  \sqrt{n_{\rm I} A M} \,, 
\end{equation} 
respectively.
For typical conditions in the NS crust $g_{\rm mix}\simeq 10^{-3}$ and $\Lambda \simeq 50 ~{\rm MeV}$.  The specific details of this calculation and possible quantum corrections to 
the sPh-lPh interaction will be reported elsewhere \cite{Reddy:2008}.
Here we employ these results to estimate $\lambda_{\rm sPh}$.

Mixing between sPhs and lPhs will damp sPh propagation because lPhs are strongly damped due to their interaction with electrons. To leading order in the small mixing parameter $g_{\rm mix}$, the mean free path of sPhs with energy $\omega$ is given by  
\begin{equation} 
\lambda_{\rm abs}(\omega)= \frac{v_s^2}{g_{\rm mix}^2}~\frac{1+ (1-\alpha^2)^2~(\omega~ \tau_{\rm lPh})^2 }
{  \alpha~(\omega~ \tau_{\rm lPh})^2}~\lambda_{\rm lPh}(\omega) \, ,
\label{eqn:mfp_abs}
\end{equation}
where $\alpha = c_s/v_s$, and $\lambda_{\rm lPh}$ and  $\tau_{\rm lPh}=\lambda_{\rm lPh}/c_s$ are the mean free path and lifetime of the lPh, respectively. At low temperature, where the Umklapp processes are frozen,  we can estimate $\lambda_{\rm lPh}$ by assuming that the lPhs decay primarily by producing particle-hole excitations (normal processes) in the degenerate electron gas. We find  that
\begin{equation} 
\lambda_{\rm lPh-N}(\omega)= f_{\rm ep}^2~\frac{2 \pi c_s^2}{p_{\rm Fe}^2~\omega}=  \frac{2}{3 \pi~c_s^2}~\frac{Z}{A} \frac{p_{\rm Fe}}{M~\omega}= \frac{2}{\pi~\omega}
\label{eqn:lPh_abs}
\end{equation}
where $f_{\rm ep} =  Z \sqrt{n_{\rm I}/A M}/c_s^2$ is related to the electron-phonon coupling constant \cite{Ziman:1960}. 

When $T \gsim T_{\rm Um}= Z^{1/3}~e^2 \omega_P/3 $,  
the Umklapp processes begin to dominate \cite{Gnedin:2001} and in this regime $\lambda_{\rm lPh-U} \simeq 100 ~a_i \simeq  ~ 2000 $ fm \cite{Chugunov:2007}. 
The corresponding  sPh mean free path is in the range $10^{-3} - 10^{-6}$ cm.  
At intermediate temperature we employ a simple interpolation formula given by
\begin{equation}
\lambda_{\rm lPh}^{-1}=f_u~ \lambda_{\rm lPh-U}^{-1} + (1-f_u)~ \lambda_{\rm lPh-N}^{-1}, 
\end{equation}
where $f_u=\exp{(-T_{\rm Um}/T)}$ \cite{Gnedin:2001}  .

Finally we estimate the mean free path due to the decay process shown in the Fig. 1 (C).  When the sPh velocity is greater than the lPh velocity this decay is kinematically allowed.
The amplitude for the sPh of momentum $q$ to split into two longitudinal lPh of momentum $q_1$ and $q_2$ is given by 
\begin{equation} 
{\cal A} \simeq \frac{g_{\rm mix}}{\Lambda^2}   ~\omega ~q_1 ~ q_2 \,.
\end{equation} 
The corresponding mean free path for a thermal phonon 
with energy $\omega = 3~T$ can be calculated and we find that 
\begin{eqnarray} 
\lambda_{\rm decay} \simeq \frac{863}{f(\alpha)}  \left(\frac{10^{-6}}{g_{\rm mix}^2}\right) \left(\frac{c_s}{0.01}\right)^7  \left(\frac{v_s}{0.1}\right)
\Lambda_0^4 ~{T_7}^{-5}~{\rm cm} 
\label{eqn:lambdasplit1}
\end{eqnarray} 
where $\Lambda_{0} = \Lambda/{\rm 50 MeV}$, $f(\alpha)=1-2 \alpha^2/3 + \alpha^4/5$, and $\alpha=c_s/v_s$. 
For the temperatures of relevance we find that $ \lambda_{\rm abs} \ll \lambda_{\rm decay} $ 
and  $ \lambda_{\rm abs} \ll\lambda_{\rm Ray}$. Consequently, only the sPh absorption process with the mean free paths given by Eq.~\ref{eqn:mfp_abs} is relevant. 

\begin{figure}[htb]
\begin{center}
\includegraphics[angle=-0,width=8.5cm, clip]{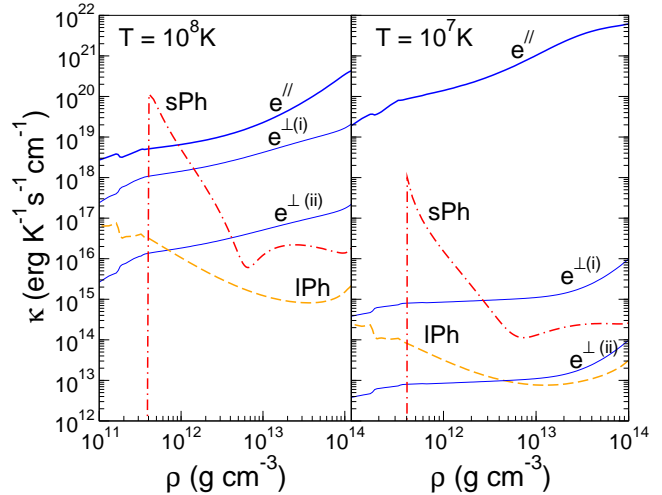}
\end{center}
\caption{Thermal conductivity in a NS crust for $T=10^{8}$~K~(left) and $T=10^7$~K~(right). The dot--dashed (red) curve shows the contribution due to sPhs.
The thick (thin) solid lines show the electron conductivity in the direction longitudinal 
(perpendicular) to the magnetic field lines
for $B=10^{13}$ G~(\rm i) and $B=10^{14}$~G~(\rm ii).
The dashed lines (yellow) is the lPh contribution.} 
\label{fig:conduct}
\end{figure}

In Fig.~\ref{fig:conduct} we employ the above result (Eq.~\ref{eqn:mfp_abs}) for the sPh mean free path and compare the 
thermal conductivity due to electron, lPhs, and sPhs transport
in a typical NS crust, at temperatures of $10^8$~K and $10^7$~K.
We show the electron contribution parallel ($e^{\parallel}$) and perpendicular 
($e^{\perp}$) to the magnetic field lines for $B=10^{13}$ G and $B=10^{14}$ G,  
according to \cite{Potekhin:1999}. 
Heat conduction due to lattice vibrations is primarily carried by the 
transverse modes because of their larger number density 
and we use the results from \cite{Chugunov:2007}.  
We can see that, above the neutron drip point, where a significant number of superfluid neutrons
coexist with the electrons and the lattice, 
heat conduction due to sPh (red dot-dashed curve) is relevant,
and is always significantly more efficient than conduction due to lPhs.  The sPh conductivity decreases with depth and reaches a minimum when the velocity of the sPh becomes equal the longitudinal speed of sound due to the resonant mixing. Near neutron drip, where $v_s \ll c_s$ sPhs can even dominate over electron conductivity along the field. In the direction transverse to magnetic field the sPh conduction is shown to be relevant over much of the inner crust for fields $B \ge 10^{14}$ G. 

\begin{figure}[htb]
\vspace{-0 cm}
\begin{center}
\includegraphics[angle=0,width=9. cm]{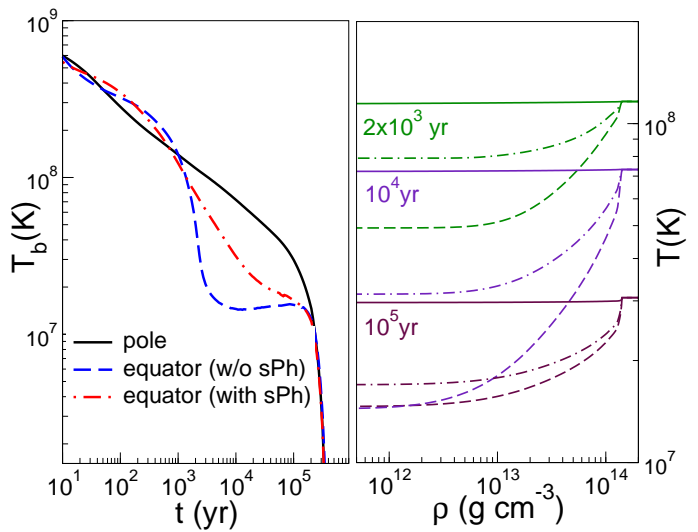}
\end{center}
\vspace{-0 cm}
\caption{Cooling curves of  magnetized NS
with and without considering sPhs. Left panel: $T_{\rm b}$ at the neutron
drip point vs. age.  Right panel: $T$ vs. density for fixed times.
The polar temperature (solid lines) is not affected, but the temperatures
at the equator with sPhs (dash-dotted lines) and without sPhs (dashes)
differ significantly.}
\label{fig:cooling}
\end{figure}

To asses how this new mode of heat conduction might affect observable aspects of NS thermal evolution we have performed magnetar cooling simulations with and without including sPh conduction. Details regarding the simulation and the input physics are described in earlier work  \cite{Aguilera:2007a}. The new ingredients here are the sPh contribution to the thermal conductivity and the use of an updated $^1S_0$ neutron superfluid gap obtained from Quantum Monte Carlo simulations \cite{Carlson:2008}. The cooling curves and temperature profiles for a NS model (Model A from Ref. \cite{Aguilera:2007a}) with a poloidal, crustal magnetic field of strength $B=5\times 10^{13}$~G at the magnetic pole, are shown in Fig.~\ref{fig:cooling}.  Temperatures at the magnetic equator are shown as dashed lines for the model without sPhs and as dashed-dotted lines for the model with sPhs. The temperature  at the pole is shown by solid lines and there is no significant difference between the two cases.  The left panel shows $T_{\rm b}$, the temperatures at the neutron drip point
($\rho \approx 3 \times 10^{11}$g~cm$^{-3}$), as a function of the NS age, $t$.
In the right panel we show the temperature profiles in the NS crust 
as a function of density, for three fixed times $t=2 \times 10^3, 10^4, 10^5$~yr. 
The effect of sPhs is important, specially in the inner crust when $T \approx 10^8$~K. In the right panel one can see that sPhs 
partially reduce the otherwise strong temperature gradient generated along the inner crust.
As a consequence, the temperature anisotropy in the inner crust is limited by sPh conduction. We note, however, that an anisotropy generated in the outer layers at density below neutron drip cannot be suppressed by sPhs. 

Finally, we remark that even the phonon conductivity can be anisotropic. Large magnetic fields will induce an anisotropy in the electron response because of the gap in the electron excitation spectrum due to Landau quantization. When this gap is much larger than the temperature, phonons propagating transverse to the magnetic field cannot excite single electron-hole states and this is likely to greatly enhance both the lPh and sPh conductivity  transverse to the field.  Rotation will create superfluid vortices that will preferentially scatter sPhs propagating perpendicular to the rotation axis. 
Both these effects warrant further investigation. 

Our estimate for the sPh conductivity, based on Eq.~\ref{conduct}, can be improved by replacing the typical mean free path by an appropriate thermal average, and by accounting for composition and nuclear structure dependence of the neutron-nucleus scattering potential for neutron-rich nuclei in the crust. Nonetheless, our study clearly demonstrates that sPhs are important for heat conduction and are primarily damped through their mixing with the longitudinal lattice phonons. The sPh contribution is significantly larger than lPhs and the magnitude of the temperature anisotropy generated at deep crustal layers is likely to be limited by superfluid heat conduction. If future observations allow us to establish a clear correlation between surface  temperature distribution, magnetic field orientation and age, we may be able to probe the superfluid nature of  the NS inner crust.  

{\it Acknowledgments:}  We thank the organizers and participants of the 
INT workshop on neutron star crust, especially, C.~Horowitz and A.~Cumming, for stimulating discussions. We also thank T.~Bhattacharya, A. Chugunov, J.~A.~Miralles and D. Yakovlev for many interesting comments.  This research
was supported by the Dept. of Energy under contract DE-AC52-06NA25396,
by the Spanish MEC grant AYA 2004-08067-C03-02 and by CONICET, Argentina. The work of SR was funded in part by the LDRD program at LANL under grant number 20080130DR. 
\vspace{-0.2in}


\begin{thebibliography}{20}
\expandafter\ifx\csname natexlab\endcsname\relax\def\natexlab#1{#1}\fi
\expandafter\ifx\csname bibnamefont\endcsname\relax
  \def\bibnamefont#1{#1}\fi
\expandafter\ifx\csname bibfnamefont\endcsname\relax
  \def\bibfnamefont#1{#1}\fi
\expandafter\ifx\csname citenamefont\endcsname\relax
  \def\citenamefont#1{#1}\fi
\expandafter\ifx\csname url\endcsname\relax
  \def\url#1{\texttt{#1}}\fi
\expandafter\ifx\csname urlprefix\endcsname\relax\def\urlprefix{URL }\fi
\providecommand{\bibinfo}[2]{#2}
\providecommand{\eprint}[2][]{\url{#2}}

\bibitem[{\citenamefont{{Yakovlev} and {Pethick}}(2004)}]{YakovlevPethick:2004}
\bibinfo{author}{\bibfnamefont{D.~G.} \bibnamefont{{Yakovlev}}}
  \bibnamefont{and} \bibinfo{author}{\bibfnamefont{C.~J.}
  \bibnamefont{{Pethick}}}, \bibinfo{journal}{\araa}
  \textbf{\bibinfo{volume}{42}}, \bibinfo{pages}{169} (\bibinfo{year}{2004}),
  \eprint{arXiv:astro-ph/0402143}.

\bibitem[{\citenamefont{Page and Reddy}(2006)}]{Page:2006ud}
\bibinfo{author}{\bibfnamefont{D.}~\bibnamefont{Page}} \bibnamefont{and}
  \bibinfo{author}{\bibfnamefont{S.}~\bibnamefont{Reddy}},
  \bibinfo{journal}{Ann. Rev. Nucl. Part. Sci.} \textbf{\bibinfo{volume}{56}},
  \bibinfo{pages}{327} (\bibinfo{year}{2006}), \eprint{astro-ph/0608360}.

\bibitem[{\citenamefont{{Haberl}}(2007)}]{Haberl2007}
\bibinfo{author}{\bibfnamefont{F.}~\bibnamefont{{Haberl}}},
  \bibinfo{journal}{\apss} \textbf{\bibinfo{volume}{308}}, \bibinfo{pages}{181}
  (\bibinfo{year}{2007}), \eprint{arXiv:astro-ph/0609066}.

\bibitem[{\citenamefont{{Geppert} et~al.}(2004)\citenamefont{{Geppert},
  {K{\"u}ker}, and {Page}}}]{Geppert2004}
\bibinfo{author}{\bibfnamefont{U.}~\bibnamefont{{Geppert}}},
  \bibinfo{author}{\bibfnamefont{M.}~\bibnamefont{{K{\"u}ker}}},
  \bibnamefont{and} \bibinfo{author}{\bibfnamefont{D.}~\bibnamefont{{Page}}},
  \bibinfo{journal}{\aap} \textbf{\bibinfo{volume}{426}}, \bibinfo{pages}{267}
  (\bibinfo{year}{2004}), \eprint{arXiv:astro-ph/0403441}.

\bibitem[{\citenamefont{Perez-Azorin et~al.}(2006)\citenamefont{Perez-Azorin,
  Miralles, and Pons}}]{Pons:2005}
\bibinfo{author}{\bibfnamefont{J.~F.} \bibnamefont{Perez-Azorin}},
  \bibinfo{author}{\bibfnamefont{J.~A.} \bibnamefont{Miralles}},
  \bibnamefont{and} \bibinfo{author}{\bibfnamefont{J.~A.} \bibnamefont{Pons}},
  \bibinfo{journal}{Astron. Astrophys.} \textbf{\bibinfo{volume}{451}},
  \bibinfo{pages}{1009} (\bibinfo{year}{2006}), \eprint{astro-ph/0510684}.

\bibitem[{\citenamefont{{Geppert} et~al.}(2006)\citenamefont{{Geppert},
  {K{\"u}ker}, and {Page}}}]{Geppert2006}
\bibinfo{author}{\bibfnamefont{U.}~\bibnamefont{{Geppert}}},
  \bibinfo{author}{\bibfnamefont{M.}~\bibnamefont{{K{\"u}ker}}},
  \bibnamefont{and} \bibinfo{author}{\bibfnamefont{D.}~\bibnamefont{{Page}}},
  \bibinfo{journal}{\aap} \textbf{\bibinfo{volume}{457}}, \bibinfo{pages}{937}
  (\bibinfo{year}{2006}), \eprint{arXiv:astro-ph/0512530}.

\bibitem[{\citenamefont{{Shternin} et~al.}(2007)\citenamefont{{Shternin},
  {Yakovlev}, {Haensel}, and {Potekhin}}}]{Shternin:2007}
\bibinfo{author}{\bibfnamefont{P.~S.} \bibnamefont{{Shternin}}},
  \bibinfo{author}{\bibfnamefont{D.~G.} \bibnamefont{{Yakovlev}}},
  \bibinfo{author}{\bibfnamefont{P.}~\bibnamefont{{Haensel}}},
  \bibnamefont{and} \bibinfo{author}{\bibfnamefont{A.~Y.}
  \bibnamefont{{Potekhin}}}, \bibinfo{journal}{\mnras}
  \textbf{\bibinfo{volume}{382}}, \bibinfo{pages}{L43} (\bibinfo{year}{2007}),
  \eprint{arXiv:0708.0086}.

\bibitem[{\citenamefont{{Brown} et~al.}(1998)\citenamefont{{Brown}, {Bildsten},
  and {Rutledge}}}]{Brown:1998}
\bibinfo{author}{\bibfnamefont{E.~F.} \bibnamefont{{Brown}}},
  \bibinfo{author}{\bibfnamefont{L.}~\bibnamefont{{Bildsten}}},
  \bibnamefont{and} \bibinfo{author}{\bibfnamefont{R.~E.}
  \bibnamefont{{Rutledge}}}, \bibinfo{journal}{\apjl}
  \textbf{\bibinfo{volume}{504}}, \bibinfo{pages}{L95+} (\bibinfo{year}{1998}),
  \eprint{arXiv:astro-ph/9807179}.

\bibitem[{\citenamefont{{Chugunov} and {Haensel}}(2007)}]{Chugunov:2007}
\bibinfo{author}{\bibfnamefont{A.~I.} \bibnamefont{{Chugunov}}}
  \bibnamefont{and}
  \bibinfo{author}{\bibfnamefont{P.}~\bibnamefont{{Haensel}}},
  \bibinfo{journal}{\mnras} \textbf{\bibinfo{volume}{381}},
  \bibinfo{pages}{1143} (\bibinfo{year}{2007}), \eprint{arXiv:0707.4614}.

\bibitem[{\citenamefont{{Shternin} and {Yakovlev}}(2006)}]{Shternin:2006}
\bibinfo{author}{\bibfnamefont{P.~S.} \bibnamefont{{Shternin}}}
  \bibnamefont{and} \bibinfo{author}{\bibfnamefont{D.~G.}
  \bibnamefont{{Yakovlev}}}, \bibinfo{journal}{\prd}
  \textbf{\bibinfo{volume}{74}}, \bibinfo{pages}{043004}
  (\bibinfo{year}{2006}), \eprint{arXiv:astro-ph/0608371}.

\bibitem[{\citenamefont{{Flowers} and {Itoh}}(1976)}]{FI76}
\bibinfo{author}{\bibfnamefont{E.}~\bibnamefont{{Flowers}}} \bibnamefont{and}
  \bibinfo{author}{\bibfnamefont{N.}~\bibnamefont{{Itoh}}},
  \bibinfo{journal}{\apj} \textbf{\bibinfo{volume}{206}}, \bibinfo{pages}{218}
  (\bibinfo{year}{1976}).

\bibitem[{\citenamefont{Son and Wingate}(2006)}]{Son:2005rv}
\bibinfo{author}{\bibfnamefont{D.~T.} \bibnamefont{Son}} \bibnamefont{and}
  \bibinfo{author}{\bibfnamefont{M.}~\bibnamefont{Wingate}},
  \bibinfo{journal}{Annals Phys.} \textbf{\bibinfo{volume}{321}},
  \bibinfo{pages}{197} (\bibinfo{year}{2006}), \eprint{cond-mat/0509786}.

\bibitem[{\citenamefont{{Baym} and {Ebner}}(1967)}]{Baym:1967}
\bibinfo{author}{\bibfnamefont{G.}~\bibnamefont{{Baym}}} \bibnamefont{and}
  \bibinfo{author}{\bibfnamefont{C.}~\bibnamefont{{Ebner}}},
  \bibinfo{journal}{Physical Review} \textbf{\bibinfo{volume}{164}},
  \bibinfo{pages}{235} (\bibinfo{year}{1967}).

\bibitem[{\citenamefont{Fetter and Walecka}(1971)}]{FetterWalecka}
\bibinfo{author}{\bibfnamefont{A.}~\bibnamefont{Fetter}} \bibnamefont{and}
  \bibinfo{author}{\bibfnamefont{J.}~\bibnamefont{Walecka}},
  \emph{\bibinfo{title}{Quantum Theory of Many-Particle Systems}},
  vol.~\bibinfo{volume}{26} (\bibinfo{publisher}{McGraw-Hill Book Co., New
  York}, \bibinfo{year}{1971}).

\bibitem[{\citenamefont{Cirigliano et~al.}(2008)\citenamefont{Cirigliano,
  Reddy, and Sharma}}]{Reddy:2008}
\bibinfo{author}{\bibfnamefont{V.}~\bibnamefont{Cirigliano}},
  \bibinfo{author}{\bibfnamefont{S.}~\bibnamefont{Reddy}}, \bibnamefont{and}
  \bibinfo{author}{\bibfnamefont{R.}~\bibnamefont{Sharma}},
  \bibinfo{journal}{in preparation}  (\bibinfo{year}{2008}).

\bibitem[{\citenamefont{Ziman}(1960)}]{Ziman:1960}
\bibinfo{author}{\bibfnamefont{J.~M.} \bibnamefont{Ziman}},
  \emph{\bibinfo{title}{Electrons and Phonons}} (\bibinfo{publisher}{Oxford
  Univ. Press}, \bibinfo{address}{Oxford}, \bibinfo{year}{1960}).

\bibitem[{\citenamefont{{Gnedin} et~al.}(2001)\citenamefont{{Gnedin},
  {Yakovlev}, and {Potekhin}}}]{Gnedin:2001}
\bibinfo{author}{\bibfnamefont{O.~Y.} \bibnamefont{{Gnedin}}},
  \bibinfo{author}{\bibfnamefont{D.~G.} \bibnamefont{{Yakovlev}}},
  \bibnamefont{and} \bibinfo{author}{\bibfnamefont{A.~Y.}
  \bibnamefont{{Potekhin}}}, \bibinfo{journal}{\mnras}
  \textbf{\bibinfo{volume}{324}}, \bibinfo{pages}{725} (\bibinfo{year}{2001}),
  \eprint{arXiv:astro-ph/0012306}.

\bibitem[{\citenamefont{{Potekhin}}(1999)}]{Potekhin:1999}
\bibinfo{author}{\bibfnamefont{A.~Y.} \bibnamefont{{Potekhin}}},
  \bibinfo{journal}{\aap} \textbf{\bibinfo{volume}{351}}, \bibinfo{pages}{787}
  (\bibinfo{year}{1999}), \eprint{arXiv:astro-ph/9909100}.

\bibitem[{\citenamefont{{Aguilera} et~al.}(2008)\citenamefont{{Aguilera},
  {Pons}, and {Miralles}}}]{Aguilera:2007a}
\bibinfo{author}{\bibfnamefont{D.~N.} \bibnamefont{{Aguilera}}},
  \bibinfo{author}{\bibfnamefont{J.~A.} \bibnamefont{{Pons}}},
  \bibnamefont{and} \bibinfo{author}{\bibfnamefont{J.~A.}
  \bibnamefont{{Miralles}}}, \bibinfo{journal}{\aap}
  \textbf{\bibinfo{volume}{486}}, \bibinfo{pages}{255} (\bibinfo{year}{2008}),
  \eprint{arXiv:0710.0854}.

\bibitem[{\citenamefont{{Gezerlis} and {Carlson}}(2008)}]{Carlson:2008}
\bibinfo{author}{\bibfnamefont{A.}~\bibnamefont{{Gezerlis}}} \bibnamefont{and}
  \bibinfo{author}{\bibfnamefont{J.}~\bibnamefont{{Carlson}}},
  \bibinfo{journal}{\prc} \textbf{\bibinfo{volume}{77}},
  \bibinfo{pages}{032801} (\bibinfo{year}{2008}), \eprint{arXiv:0711.3006}.

\end{thebibliography}


\end{document}